\documentclass[12pt]{article}

\newcommand{\be}{\begin{equation}}
\newcommand{\ee}{\end{equation}}
\newcommand{\bqn}{\begin{eqnarray}}
\newcommand{\eqn}{\end{eqnarray}}

\newcommand{\fr}{\frac}
\newcommand{\lb}{\label}

\newcommand{\om}{\omega}

\newcommand{\te}{\theta}

\newcommand{\del}{\partial}

\newcommand{\de}{\delta}

\newcommand{\st}{\star}
\newcommand{\ti}{\tilde}
\newcommand{\da}{\dagger}

\begin{document}

\begin{flushright}
hep-th/0202062
\end{flushright}

\vspace{13mm}
\begin{center}
{\Large\bf Wigner Functions for the Landau Problem in  Noncommutative Spaces}

\vspace{17mm}

\"{O}mer F. Dayi$^{a,b,}$ 
\footnote{E-mail: dayi@itu.edu.tr and dayi@gursey.gov.tr.}
\,{and}\,
Lara T. Kelleyane$^{a,}$ 
\footnote{E-mail: kelleyane@itu.edu.tr.}\\
\vspace{5mm}

\end{center}

\noindent
{\em $^{a}${\it Physics Department, Faculty of Science and
Letters, Istanbul Technical University,\\
80626 Maslak--Istanbul, Turkey.} } 

\vspace{3mm}

\noindent
{\em $^{b}${\it Feza G\"{u}rsey Institute, P.O. Box 6, 81220,
\c{C}engelk\"{o}y, Istanbul, Turkey. } }

\vspace{1cm}

\begin{center}
{\bf Abstract}
\end{center}

\vspace{3mm}

%\begin{abstract}
{\small

An electron moving on plane in a uniform
magnetic field orthogonal to plane
  is known as the
Landau problem.  Wigner  functions
for the Landau problem when 
the plane is noncommutative are found 
employing
solutions of the Schr\"{o}dinger equation 
as well as solving the ordinary $\st$--genvalue
equation in terms of an effective Hamiltonian.
Then, we let momenta and coordinates of the
phase space be noncommutative
and introduce a generalized $\st$--genvalue equation. 
We solve this equation to find
the related  Wigner functions and show  that
under an appropriate  choice of noncommutativity relations
they are independent of noncommutativity parameter.  }

%\end{abstract}

%\end{titlepage}   

\newpage

\section{Introduction}

Deformation quantization 
which is also known as
Weyl--Wigner--Groenewold--Moyal quantization
describes a quantum system in terms of
c--number phase space variables\cite{wwgm}. The correspondence with the
operator quantization is  due to symbol maps and
star products. Operators are mapped into c--number functions
however, their composition is given by star product which 
is noncommutative but associative. 
One of the most important ingredients of
deformation quantization is Wigner function\cite{wig}.
It is essential to define phase space integrals
corresponding to expectation values of operator
formalism.

When classical coordinates are converted into
 noncommuting variables
they mutate into operators. 
Thus after performing canonical quantization
of noncommutative systems one should
deal with two kind of operators. This can be avoided 
by formulating noncommutativity of coordinates as  deformation
of commuting ones. 
One can adopt all the machinery developed
in deformation quantization.

An electron moving on plane in a uniform external
magnetic field orthogonal to plane is known as the
Landau problem.  Different aspects of the Landau
problem in noncommutative coordinates 
have been widely 
studied\cite{all}. Our aim is to derive 
Wigner functions for the Landau problem 
for two different noncommuting spaces: 
When only  coordinates are noncommuting and then
letting  coordinates as well as  momenta are noncommuting.

In Section 2 we briefly recall definitions  of  Wigner 
function. In Section 3 we 
study the Landau problem 
on the  noncommutative plane 
by adopting 
a differential representation for  
quantum mechanical operators.
This yields Wigner functions which
are formal. However, guided by this formulation 
we define an effective Hamiltonian and
obtain  Wigner functions utilizing 
the usual deformation quantization.
When both  quantum mechanics and noncommutativity
are formulated  as deformations,
it is plausible to consider all of phase space variables 
on the same footing. Indeed, in Section 4 deformation quantization
is adopted by letting coordinates as well as momenta are noncommuting
for the Landau problem and the related  Wigner functions are
calculated. Moreover, we show that  noncommutativity relations 
can be  chosen  such that the resultant Wigner functions
are not aware of noncommutativity of phase space variables.
These Wigner functions can be used to calculate some 
physical quantities in noncommutative spaces.

\section{Wigner Functions }

Wigner functions can be defined equivalently in two different ways.
One of them 
refers to the operator formulation of  quantum
mechanics:  Let the Hamiltonian operator  be $\hat{\cal H}$ and define 
the eigenvalue problem  
\be
\lb{ei}
\hat{\cal H} \psi (\vec{x}) ={\cal E} \psi (\vec{x}) .
\ee
If we denote its spectrum by ${\cal E}_l$ and the corresponding eigenstates
by $\psi_l(\vec{x}),$  (diagonal) Wigner functions
are defined as
\be
\lb{qd}
W_l(\vec{x},\vec{\pi})=\frac{1}{2\pi }\int d^ds\  \psi_l^*(\vec{x}-{\hbar 
\over 2}\vec{s})
e^{-i\vec{s}\cdot \vec{\pi}}\psi_l (\vec{x}+{\hbar \over 2}\vec{s}),
\ee
where  $(\vec{x},\vec{\pi})$ are $2d$ dimensional phase space variables.

Wigner functions can also be defined without referring 
to the operator formalism of quantum mechanics (\ref{ei}), by introducing
the $\hbar$--star product
\be
\label{sh}
\st_\hbar \equiv\exp {i\hbar\over 2} \Big(
{\stackrel\leftarrow\del }_{\vec{x}} \cdot
{\stackrel\rightarrow\del}_{\vec{\pi}}
-{\stackrel\leftarrow\del}_{\vec{\pi}} \cdot
{\stackrel\rightarrow\del}_{\vec{x}}
\Big) .
\ee 
Now, the  Wigner functions are  defined as $\st_\hbar$--genvalue equation
\be
\lb{sgv}
{\cal H}\st_\hbar W_l(\vec{x},\vec{\pi})={\cal E}_lW_l(\vec{x},\vec{\pi}),
\ee
where $\cal H$ is the classical Hamiltonian function
corresponding to $\hat{\cal H}.$

These two definitions are equivalent\cite{wig}.
Wigner function are used to define integrals 
in the deformation quantization
corresponding to expectation values in
operator formulation.

\section{Wigner Functions for the Landau Problem
 in Noncommutative Coordinates}

An electron moving in the plane which is
perpendicular to the  external uniform magnetic field $B,$
neglecting spin, is 
described by the Hamiltonian
\be
\label{fh}
H = {1\over 2m}({\vec p}+{e\over c}{\vec A})^2.
\ee
where we adopt the symmetric gauge
\be
\label{gco}
{\vec A}=(-{B\over 2}y,{B\over 2}x).
\ee

Consider  the coordinates satisfying
\be
\label{com}
[\tilde{x},\tilde{y}]=i\te,
\ee 
where $\te$ is a constant parameter.
Because of being 
operators even at the classical level,
we denoted noncommuting coordinates as $\ti{x},\ti{y}.$ 
Noncommutativity can be imposed 
by treating the coordinates as
commuting, but
introducing the $\te$--star product 
\be
\label{star}
\st_\te \equiv\exp \frac{i\te}{2} \Big(
{\stackrel\leftarrow\del }_{x} {\stackrel\rightarrow\del}_{y}
-{\stackrel\leftarrow\del}_{y}{\stackrel\rightarrow\del}_{x}
\Big) .
\ee 
Now, we deal with  the commutative coordinates $x$ and $y$ but 
composition of functions is given by 
$\te$--star product 
(\ref{star}). For example, 
instead of the commutator  (\ref{com}) one introduces the Moyal bracket
\be
x\st_\te y-y\st_\te x=i\te .
\ee 
We quantize  this system in terms of the standard canonical quantization
by establishing the usual canonical commutation relations
\be
[r_i,\hat{p}_j]=i\hbar \delta_{ij}
\ee
and adopting the usual differential representation
\be
\lb{dre}
 \hat{p}_i=-i\hbar {\del\over \de r_i}.
\ee
Note that we use the same notation for  the  
(quantum) operator and classical coordinates $\vec{r}=(x,y)$.

One of the receipts to define quantum mechanics
in noncommutative coordinates is to interpret
derivatives as momentum operators: $\vec{\del}=\frac{i}{\hbar}\hat{\vec{p}}.$
According to this receipt,  
in the symmetric gauge (\ref{gco})
the Hamiltonian operator $\hat{H}$ corresponding to (\ref{fh}), acting 
on an arbitrary function  $\Psi (\vec{r})$ yields
\be
\hat{H} \st_\te \Psi (\vec{r}) =
\frac{1}{2m}\left[
\left( \hat{p}_x -\frac{eB}{2c} y \right)^2 +
\left( \hat{p}_y +\frac{eB}{2c} x \right)^2 \right]\st_\te \Psi (\vec{r})
\equiv \hat{H}_{nc} \Psi (\vec{r}),
\ee
where, in terms of $\kappa =-\frac{eB\theta}{4c\hbar },$
we  defined
\be
\hat{H}_{nc} =
\frac{1}{2m}\left[
\left( (1+\kappa )\hat{p}_x -\frac{eB}{2c} y \right)^2 +
\left( (1+\kappa )\hat{p}_y +\frac{eB}{2c} x \right)^2 \right].
\ee
The eigenvalue equation
\[
\hat{H}_{nc}\phi(\vec{r})=E \phi (\vec{r}) ,
\]
can be solved in polar coordinates
and its
eigenstates  
can be  given in terms of the Laguerre polynomials in $(x^2+y^2).$
These solutions are not suitable to
find the related Wigner functions 
although, formally we can write them
by making use of the definition (\ref{qd}).
Another method may be to use 
definition of Wigner functions in polar coordinates\cite{rad}.
 
We would like to use the fact that effectively the Landau problem 
is one dimensional harmonic oscillator. The operator
\be
\label{oa}
a= \frac{1}{\sqrt{2m \hbar \ti{\omega}}}\left[
\left( \hat{p}_x-\frac{m\ti{\omega}}{2} y\right) 
-i\left( \hat{p}_y+\frac{m\ti{\omega}}{2} x\right) 
\right] ,
\ee
and its hermitian conjugate $a^\dagger$ 
can be shown to satisfy
\be
[a,a^{\da}]=1,
\ee
where $\ti{\om}\equiv  (1+\kappa ) \omega$ 
and $\om={eB\over mc}$ is the cyclotron frequency.

The Hamiltonian becomes
\be
\label{ham}
\hat{H} = \hbar\ti{\om}(a^{\da}a+{1\over 2}),
\ee
which is equivalent to harmonic oscillator in one dimension
whose frequency is $\ti{\om}.$
Thus, its spectrum is given by 
\[
E_n=\hbar \ti{\om}(n+1/2),
\]
where $n=0,1,\cdots .$
In terms of the ground state
\[
 \phi_0 (\vec{r},\te )={m\ti{\om}\over 2\pi \hbar}\exp[{-m\ti{\om}\over 4\hbar} (x^2+y^2)],
\]
its eigenfunctions are given
as
\[
\phi_n(\vec{r},\te )= \fr{1}{\sqrt{n !}}(a^\dagger)^n\phi_0.
\]
These states  can be used to write the 
Wigner functions as
\[
W_{L,n}(\vec{r},\vec{p},\te )=
{1\over 4\pi^2}\int d^2v\  \phi_n^*(\vec{r}-{\hbar \over 2}\vec{v})
e^{-i\vec{v}\cdot \vec{p}}
\phi_n(\vec{r}+{\hbar \over 2}\vec{v}).
\]
However, as it is mentioned above these Wigner
functions are only formal, few of the integrals 
can be calculated under some  conditions.

Although we are using the specific representation (\ref{dre}),
let us assume that we can choose the realization
\be
\lb{wr}
a\psi(q)={1\over \sqrt{2}}(q+{\del \over \del q}) \psi(q),\  
a^\dagger\psi (q)={1\over \sqrt{2}}(q-{\del \over \del q})\psi(q)  ,
\ee
where the parameter $q$ is a c--number. 
Then, eigenfunctions of (\ref{ham}) coincide with the
eigenfunctions of one dimensional
harmonic oscillator, which lead to
Wigner functions\cite{all}
\be
W_{0,n}(P,q)=\frac{(-1)^n}{\pi \hbar}e^{-z}L_n(2z) ,
\ee
where we used the variable
\[
z=(P^2+q^2).
\]
$P$ is the parameter appearing in the definition (\ref{qd}).

As we have already mentioned in Section 2, there exists
another method of calculating Wigner functions:
We can use the $\hbar$--star product (\ref{sh}) and the
definition (\ref{sgv}). 
However, in this 
case we should define a c--number object corresponding to
$\hat{H}_{nc}.$ Let us define the effective c--number Hamiltonian
\be
\lb{hef}
H_{eff}=\frac{(1+\kappa)^2}{2m}\ \vec{p}\ ^2+\frac{m\om^2}{8}{\vec{r}}\ ^2
+\frac{\ti{\om}}{2}(xp_y-yp_x).
\ee
We would like to emphasize the fact
that it is not classical
because  $\kappa$ depends on $\hbar .$
Now, to find the related 
Wigner functions 
we use the definition
\be
H_{eff}\st_\hbar W_1(\vec{p},\vec{r},\te )=EW_1(\vec{p},\vec{r},\te ).
\ee
This can be written as a differential equation only in one variable
\[
\rho \equiv H_{eff}
\]
as
\be
\lb{er}
\rho \fr{d^2 W_1(\rho)}{d\rho^2}
+ \fr{d W_1(\rho)}{d\rho} -\fr{4}{\hbar^2\ti{\om}^2}(\rho +E)  W_1(\rho )=0.
\ee
To solve this equation let us change the variable as
\be
R=\fr{4\rho}{\hbar \ti{\om}}
\ee
and write
\be
W_1=e^{-R/2}L(R).
\ee
Thus (\ref{er}) becomes
\be
\left[ R \frac{d^2}{dR^2} +(1-R) \frac{d}{dR}-(\frac{1}{2}-
\frac{E}{\hbar\ti{\om}})\right] W_1=0,
\ee
whose solutions are Laguerre polynomials $L_n(4R)$ where $n=0,1,\cdots $ and
\[
E_n=\hbar \ti{\om}(n+1/2).
\]
Thus we have
\be
W_{1,n}(\vec{p},\vec{r},\te )=e^{-2H_{eff}/\hbar\ti{\om}}L_n(4H_{eff}/\hbar \ti{\om} ).
\ee
Indeed, this is not surprising: The system is equivalent
to one dimensional harmonic oscillator, which
also can be observed  at the ``classical level".
By performing the canonical transformation
\bqn
p_1={(1+\kappa)\over \sqrt{m}}p_x -{\om \sqrt{m} \over 2}y, \  
q_1={(1+\kappa)\over \sqrt{m}}p_y +{\om \sqrt{m} \over 2}x,  \\ 
p_2={(1+\kappa)\over \sqrt{m}}p_y -{\om \sqrt{m} \over 2}x,  \
q_2={(1+\kappa)\over \sqrt{m}}p_x +{\om \sqrt{m} \over 2}y  ,
\eqn
the effective Hamiltonian becomes
\be
H_{eff}={1 \over 2}(p_1^2+q_1^2).
\ee
The other set of variables $(p_2,q_2)$ leads to the degeneracy of the Landau levels
after quantization.

As it may be expected, using the effective Hamiltonian
(\ref{hef}) and assuming the representation (\ref{wr}) yield the 
same Wigner functions.

\section{Wigner Functions for the Landau Problem 
in Noncommutative Phase Space }

Quantum mechanics in noncommutative space can be defined 
in terms of Wigner functions found
by making use of the $\st_\hbar$ and $\st_\te$ simultaneously.
However, when $\st_\hbar$--genvalue equations are considered
it is plausible to treat
$\vec{r}$ and $\vec{p}$ 
on the same footing. 
Thus, we would like to 
deform also momenta of the phase space by introducing
\be
\label{starp}
\st_{k}\equiv\exp \frac{ ik\te}{2} \Big(
{\stackrel\leftarrow \del }_{p_x} {\stackrel\rightarrow\del }_{p_y}
-{\stackrel\leftarrow\del}_{p_y}{\stackrel\rightarrow\del }_{p_x}
\Big) ,
\ee 
where $k$ is any constant.

Now, consider
\[
\st \equiv  \star_\hbar \star_\theta \star_{\theta^\pm} ,
\]
which we use to define
the $\st$--genvalue equation
\be
H \star \Psi = E\Psi ,
\ee
where $H$ is the classical Hamiltonian (\ref{fh})
of the Landau problem in the symmetric gauge (\ref{gco}). 

A similar approach for harmonic oscillator was given in \cite{hs}.

$\st$--genvalue equation can be written as
\bqn
%\begin{array}{c}
& \{ H - \fr{\hbar^2}{8}  [ \fr{1}{m}
\vec{\del}_r \cdot\vec{\del}_r 
+ 
\fr{m\om^2}{4}
\vec{\del}_p \cdot \vec{\del}_p
-\om ( \del_{p_x}\del_y - \del_{p_y}\del_x )] & \nonumber \\
&-\fr{\theta^2}{8} [ \fr{m\om^2}{4}\vec{\del}_r \cdot\vec{\del}_r 
 + {k^2\over m}\vec{\del}_p \cdot \vec{\del}_p
 +k \om 
( \del_{p_x}\del_y - \del_{p_y}\del_x ]  & \nonumber \\ 
& +\fr{\theta \hbar  }{4} [{\om \over 2}\vec{\del}_r \cdot \vec{\del}_r 
+k \vec{\del}_p \cdot \vec{\del}_p
 +({\om^2m \over 4} + {k \over m})
( \del_x \del_{p_y} - \del_y \del_{p_x} ) & \label{den} \\
& - \frac{i\hbar}{2} [{1 \over m} \vec{p}\cdot \vec{\del}_r  - {m\om^2 \over 
4}\vec{r} \cdot \vec{\del}_p 
-{\om \over 2}( y_x \del_x - p_x \del_{p_y} - x \del_y + 
p_y \del_{p_x} ) ]   & \nonumber \\
& +\fr{i\theta}{2} [  {m\om^2 \over 4}(x \del_y - y \del_x )
+ {k \over m}( p_x \del_{p_y}  - p_y \del_{p_x} ) 
 + \fr{ \om }{2} ( p_y \del_y  + y \del_{p_y} -k  x \del_{p_x} 
-k p_x \del_x )] -E \} W_2 =0 , &  \nonumber
\eqn
where $\vec{\del}_r \equiv \del / \del \vec{r}$
and $\vec{\del}_p \equiv \del / \del \vec{p}.$
Introducing the variable
\be 
\zeta \equiv H,
\ee
one can show that the imaginary part of (\ref{den}) vanishes and 
it reads
\be  
\lb{bi}
-4\alpha^2 (k, \te , \hbar )  
(\zeta \fr{d^2 W_2(\zeta)}{d \zeta^2}
+\fr{dW_2(\zeta)}{d\zeta} ) 
+ \zeta W_2(\zeta) = E W_2(\zeta) ,
\ee
where we defined
\be
\lb{apm}
\alpha^2 (k,\te,\hbar )={1\over 16} \hbar^2\om^2 + {\te^2\over 16}
(\fr{k^2}{m^2}+{\om^4m^2 \over 16}+ {k\om^2 \over 2} )
-{1\over 2}\hbar \te ({\om^3m\over 4} +{k \om \over m}).
\ee
In terms of the variable
\[
\eta ={\zeta \over \
\alpha (k,\te,\hbar )},
\]
(\ref{bi}) can be written as
\[
(\eta{\del_\eta}^2 +\del_\eta -\eta/4 +\ti{E})W_2(\eta)=0,
\]
where $\ti{E}={E\over 4\alpha (k,\te ,\hbar )}.$
By writing
\[
W_2(\eta )=e^{-\eta /2}L(\eta),
\]
we see that $L(\eta )$ satisfies the differential equation
\[
[\eta {\del_\eta}^2+(1-\eta)\del_{\eta} +\ti{E}-1/2]L(\eta)=0.
\]
Thus $L(\eta)$ is given as Laguerre polynomials $L(\eta)=L_n(4\eta)$ and
\[
E_{k ,n}=4\alpha (k,\te ,\hbar ) (n+1/2),
\]
where as usual $n=0,1,\cdots .$
Therefore, Wigner functions of the Landau problem in
noncommutative phase space are
\be
\lb{lw}
W_{2,n}(\vec{p},\vec{r}, k,\te)= \fr{(-1)^n}{\pi\hbar} e^{-H/2\alpha (k,\te ,\hbar )}
L_n(4H/\alpha (k,\te ,\hbar )).
\ee

There exists a very interesting case: When
\[
k\equiv k_s=-\fr{1}{4}\om^2m^2
\]
$\alpha (k,\te ,\hbar )$ becomes 
independent of $\te :$
\[
\alpha^2 (k_s ,\te , \hbar )= \fr{\hbar^2 \om^2}{16}
\]
Thus, noncommutativity relations of phase space can be taken
such that the resultant quantum mechanics is not aware of 
noncommutativity.

One can use the Wigner functions (\ref{lw}) to calculate
physical quantities. One of the first applications is to 
calculate the quantities  considered in
other frameworks\cite{all} and compare the results.
Nevertheless, this approach constitutes  an alternative 
method of defining the Landau problem in noncommutative 
classical phase space.

\pagebreak

\end{document}